\documentclass[A4paper,12pt]{article}
\usepackage{amsfonts}
\usepackage{amsmath}
\usepackage{amssymb}
\usepackage{graphicx}
\begin{document}
\title{Parasupersymmetric Quantum Mechanics of Order $3$ and a Generalized Witten Index}
\author{Marko Sto\v si\'c and Roger Picken \\
Departamento de Matem\'{a}tica
and\\
CEMAT -
Centro de Matem\'{a}tica e Aplica\c{c}\~{o}es\\
Instituto
Superior T\'{e}cnico\\ Av. Rovisco Pais\\ 1049-001 Lisboa\\ Portugal\\
e-mail: mstosic@math.ist.utl.pt \quad rpicken@math.ist.utl.pt}
\date{}
\maketitle
\begin{abstract}
In this paper we generalize the Rubakov-Spiridonov
parasupersymmetry algebra to the order 3 case. We also generalize the notion of the Witten index, 
and we provide a class of models satisfying our parasupersymmetry algebra. Finally, we show that there is a correspondence between the Hamiltonian and the index in our class of models.
\end{abstract}

\newtheorem{theorem}{Theorem}
\newtheorem{proposition}{Proposition}
\newtheorem{definition}{Definition}
\newtheorem{remark}{Remark}

\newcommand{\ud}{\mathrm{d}}

\def\kraj{\hfill\rule{6pt}{6pt}}
\def\diag{\mathop{\rm diag}}
\def\deg{\mathop{\rm deg}}
\def\rank{\mathop{\rm rank}}
\def\sgn{\mathop{\rm sgn}}
\def\trace{\mathop{\rm Tr}}
\def\ind{\mathop{\rm ind}}
\def\ker{\mathop{\rm ker}}

\def\diag{\mathop{\rm diag}}
\def\and{\mathop{\rm and}}
\def\F{\mathcal{F}}
\def\R{\mathbb{R}}
\def\C{\mathbb{C}}
\def\H{\mathcal{H}}
\def\P{\mathcal{P}}
\def\I{\mathbb{I}}
\def\Q{Q^{\dagger}}
\def\m{Q^2{Q^{\dagger 2}}}
\def\n{{Q^{\dagger 2}}Q^2}
\def\a{Q{Q^{\dagger}}} 
\def\b{Q^{\dagger}Q}
\def\K{{Q^{\dagger 2}}}
\def\T{{Q^{\dagger 3}}}
\arraycolsep 2pt
\maketitle
\newpage
\section{Introduction}

\indent The notion of supersymmetry - symmetry between bosonic and fermionic degrees of freedom - is an idea that is 
believed to play an important role in physics. For an excellent review, see \cite{GK}. However, it is not clear that nature is supersymmetric at the level of elementary particles.  
Besides the ordinary Bose-Einstein and Fermi-Dirac statistics there exists
an intermediate set of para-Fermi generalizations
\cite{OK}, characterized by an integer $p$, indicating the maximum number of particles which may occupy the same state. Ordinary fermions correspond to $p=1$ and bosons to $p=\infty$.
The particular case
$p=3$ is especially interesting since it is an alternative way of describing quarks (without using colours) - see \cite{G}.
The natural generalization of supersymmetry - so-called
parasupersymmetry (abbreviated to parasusy) - is a symmetry between bosonic and parafermionic degrees of freedom (more precisely, between ``neighbouring'' parafermion levels, taking the bosonic level to be parafermion level 0), and one might hope that it 
plays a similarly important role to supersymmetry in the description of nature.
The parasupersymmetric structure will of course depend on the number,
$p$, of parafermions that can occupy the same state. \\
\indent Supersymmetric quantum mechanics, the $p=1$  case, was studied by Witten in \cite{W}. The generalization of this to the $p=2$ case
was done by Rubakov and Spiridonov in \cite{RS}, and  also in a slightly different way by Beckers and Debergh in \cite{BD}. In this paper we present the
generalization of these results to the $p=3$ case. We give a natural form of the  parasupersymmetry algebra and present a class of models satisfying
it. Also, we will discuss the generalization proposed in \cite{P} of the Witten index \cite{W} in this case, and relate the index to the spectra of
the Hamiltonian in our $p=3$ model, as well as describing the same correspondence in the cases $p=1$ and $p=2$.\\
\indent The paper is organized as follows: in section 2 we recall some basic features of the $p=1$ and $p=2$ cases. In section 3 we give the definition of the generalized $p=3$ parasupersymmetry algebra, together with the 
definition of generic families and the stratification of the Hilbert space into a direct sum of certain low-dimensional
invariant subspaces. In section 4 we recall the definition \cite{P} of the parasupersymmetry index for $p=3$, generalizing the Witten index, and prove a formula for its calculation. In section 5, we generalize the class of models
of section 2 to the $p=3$ case, and prove that it satisfies the relations of our parasusy algebra from section 3. Finally, in
section 6, we prove that the spectrum of the Hamiltonian of our model is generically 4-fold degenerate and positive, except
possibly at the lowest (vacuum) levels, and furthermore, that there is a correspondence between the index of the model and the
possible forms of the spectrum. We also present the same results (correspondence between the index and the spectrum) for the models
in the cases $p=1$ and $p=2$.

\section{The Cases $p=1$ and $p=2$}   

\indent Let us review briefly the supersymmetric case $p=1$. We will denote the bosonic coordinate by $x$, and the operator of taking the $x$-derivative by $d_x$ (we avoid using the momentum operator explicitly, so as not to have to introduce the imaginary unit $i$ into the expressions). One can introduce supersymmetry operators, also known as supercharges, by:
\begin{equation}
Q=\left( \begin{array}{cc} 0 & 0 \\ -d_x+W(x) & 0 \end{array} \right),\quad 
\Q=\left( \begin{array}{cc} 0 & d_x+W(x) \\ 0 & 0 \end{array} \right),
\end{equation}
where $W(x)$ is an arbitrary function 
(of class ${\cal C}^2$), 
known as the superpotential. The supercharges obey the following algebra:
\begin{eqnarray}
&Q^2=\K=[H,Q]=[H,\Q]=0,&  \nonumber \\
&\a+\b=2H,&
\end{eqnarray}
where the Hamiltonian is given by:
$$H=\frac{1}{2}(-d_x^2+W^2)\left( \begin{array}{cc}1&0\\0&1 \end{array} \right)+
\frac{1}{2}\left( \begin{array}{cc} W'&0\\0&-W' \end{array} \right).$$
The Hamiltonian and (conserved) supercharges act on the Hilbert space 
$$\H=L^2(\R,{\R}^2)=\left\{ \left( \begin{array}{c} f_1\\f_0 \end{array} \right) | f_i \in L^2(\R,\R) \right\}$$ 
with the natural inner product coming from the $L^2$ inner product and the standard inner product on ${\R}^2$.
If we define $\H_1$ to be the subspace with $f_0=0$ and $\H_0$ to be the subspace with $f_1=0$, then we have the following
stratification:
\begin{eqnarray*}
\H=\H_0 \oplus \H_1, &\,\,& H(\H_i) \subset \H_i, \\ 
Q(\H_1) \subset \H_0,&\,\,& Q(\H_0)=\{0\},\\  
\Q(\H_0) \subset \H_1,&\,\,& \Q(\H_1)=\{0\}.
\end{eqnarray*}
\indent Following \cite{RS} the simplest generalization of (1) to the $p=2$ case is given by the following parasusy operators:
\arraycolsep 2pt
\begin{equation}
Q=\left( \begin{array}{ccc} 0 & 0 & 0 \\ -d_x+W_1 & 0 & 0 \\ 0 & -d_x+W_2 & 0 \end{array} \right),\quad 
\Q=\left( \begin{array}{ccc} 0 & d_x+W_1 & 0\\ 0 & 0 & d_x+W_2\\ 0 & 0 & 0\end{array} \right),
\end{equation}
where $W_1(x)$ and $W_2(x)$ are functions (of class ${\cal C}^2$) whose properties are still to be specified. We only have to find the Hamiltonian
$H$ in order to obtain a suitable algebra generated by $H$, $Q$ and $\Q$. This time, bilinear combinations of $Q$ and $\Q$
cannot have the form $-d_x^2 {\I}_3 + U(x)$ with $U(x)$ a $3 \times 3$ matrix depending only on the coordinate $x$.
However, the trilinear combinations satisfy:
\begin{eqnarray}
&Q^3=\T=[H,Q]=[H,\Q]=0,&  \nonumber \\
&Q^2\Q+Q\b+\Q Q^2=4QH,&\\
&\K Q+\Q \a+Q \K= 4 \Q H,& \nonumber \\ \nonumber
\end{eqnarray} 
where the Hamiltonian is given by:
\begin{equation*}
4H = (-2d_x^2+W_1^2+W_2^2){\I}_3+\diag (3W_1'+W_2',-W_1'+W_2',-W_1'-3W_2'),
\end{equation*}
provided that the superpotentials satisfy:
\begin{equation}
{(W_2^2-W_1^2)'+{W_2}''+{W_1}''}=0.
\end{equation}
Here all the operators are acting on the Hilbert space $\H=L^2(\R,{\R}^3)$ with its canonical inner product, and we have a stratification:
\begin{eqnarray*}
\H=\H_0 \oplus \H_1 \oplus \H_2, &\,\,& H(\H_j) \subset \H_j, \\
Q(\H_2) \subset \H_1,\,\,Q(\H_1) \subset \H_0,&\,\,& Q(\H_0)=\{0\},\\
\Q(\H_0) \subset \H_1,\,\,\Q(\H_1) \subset \H_2,&\,\,& \Q(\H_2)=\{0\}.
\end{eqnarray*} 
In both cases $p=1$ and $p=2$, we can introduce the (para)fermion number operator $N$, which on each $\H_j$ is given by multiplication by $j$ and
on the rest of $\H$ by linear extension. In the supersymmetric case it coincides with the fermion number operator $F$ used in \cite{W}.\\
\indent Now, even if we forget about the specific forms (1) and (3) and work just with the respective parasupersymmetry algebras (2) and (4),
we can find further structures and a finer stratification of the Hilbert space. For the case $p=1$ this was analyzed in
\cite{W} and for the case $p=2$ in \cite{P}. In both cases it can be shown that the operators $H, \a, \b$ and $N$ commute, and that, if $\Psi$ is an eigenstate
of all four operators then so are $Q \Psi$ and $\Q \Psi$. Thus we can define the family $\F(\Psi)$ corresponding to such
a state, as being the subspace spanned by $\Psi$ and all the states that can be obtained from $\Psi$ by acting on it with $Q$'s and
$\Q$'s. Such a family will be an invariant subspace for each of the operators $H, Q, \Q$ and $N$. Also, the dimension
of each such family is less than or equal to $p+1$, and if it is equal to $p+1$  we will call such a family generic. Finally, with 
suitable restrictions on the Hamiltonian (having a discrete spectrum, being bounded below and with a finite multiplicity for each
eigenvalue) it can be proved that there exists a sequence $\Psi_n \in \H$ such that:
$$\H=\bigoplus_{n=1}^{\infty} \F(\Psi_n).$$
Further details for the cases $p=1$ and $p=2$ can be found in \cite{W} and \cite{P}.\\
\indent In the next section we will
introduce a natural generalization of the algebras (2) and (4) to the $p=3$ case and prove the analogues of all the above statements. 

\section{The $p=3$ Parasupersymmetry Algebra}

As indicated in the previous section, the algebraic relations (2) and (4) satisfied by the operators give rise to a decomposition of the Hilbert space into natural finite-dimensional subspaces. This statement holds irrespective of any particular model obeying the algebra. We will now develop the same algebraic perspective for the case $p=3$, starting with the following definition.

\begin{definition}Let $\H$ be a separable Hilbert space with inner product 
$\langle\,  , \, \rangle$ and $H$ a self-adjoint Hamiltonian
operator on
$\H$ with purely discrete spectrum.
$(\H,H)$ will be said to be {\em parasupersymmetric of order $p=3$} iff the following three conditions hold:\\
1.\quad $\H=\H_0\oplus \H_1\oplus \H_2\oplus \H_3,\quad H(\H_j)\subset \H_j$\\
2.\quad There exists a densely defined operator Q on $\H$, with adjoint $\Q $ such that:
\begin{center}
$Q(\H_0)=0,\quad Q(\H_j)\subset \H_{j-1},\quad j=1,2,3$
\end{center}
\begin{center}
$\Q(\H_3)=0,\quad \Q(\H_{j-1})\subset \H_j,\quad j=1,2,3$
\end{center}
3.\quad The operators H, Q, $\Q$ satisfy the following parasusy algebra:
\begin{eqnarray*}
(i)&[Q,H]=[\Q,H]=0\\
(ii)&Q^3\Q +Q^2\Q Q+Q\Q Q^2+\Q Q^3=6Q^2H\\
(iii)&Q\K Q=\Q Q^2 \Q\\
(iv)&\T Q+\K Q\Q +{\Q}Q{\Q}^2 + Q\T =6\K H
\end{eqnarray*}
\end{definition}
\begin{remark} This algebra generalizes the Rubakov and Spiridonov \cite{RS} algebra of equation (4), and it is partly of the form
suggested in \cite{P}. Relation $(iv)$ is included just for the sake of symmetry since it is the adjoint of $(ii)$.
Also, relation $(ii)$ is symmetric in the following sense: we have all four
quadrilinear combinations with one 
$Q$ and three $\Q$'s with the same coefficient (equal to 1) on the left hand side, 
and the right hand side has the same homogeneity as the left hand side if $H$ is replaced by $\a$ or $\b$, analogously to the $p=1$ and $p=2$ algebras (2) and (4). 
The constant on the right hand side of relations $(ii)$ and $(iv)$ is arbitrary, but we choose it to be $6$, since for the model in section 5 this choice leads to a Hamiltonian with the standard factor $-1/2$ multiplying $d_x^2$ in the kinetic term. Also, for the decomposition into families that we prove below, we need the operators $\a$, $\b$ to commute, i.e. relation $(iii)$. Note, that this is an expression involving two quadrilinear combinations with
two $Q$'s and two $\Q$'s, but is much simpler than the form suggested in \cite{P}. Thus the relations $(i)-(iv)$ are, in fact, the ``minimal'' set of
conditions required in order to have the fine stratification properties that we are looking for. Finally, the algebra is satisfied by the natural $p=3$
generalization of the class of models from section 2, as will be shown in section 5. 
\end{remark}
 
\indent The parafermion number operator $N$ is given on the ${\H}_j$'s by:
$$N(f)=jf,\quad f\in\H_j,$$
and on the whole of $\H$ by linear extension.
\\
\begin{proposition}
The operators H, Q$\Q$, $\Q$Q, N, $\m$, $\n$, $Q\K Q$, $\Q Q^2 \Q$, $Q\Q Q\Q$ and $\Q Q\Q Q$  commute.
\end{proposition}
\indent
\textit{Proof}: Since the last four operators are products of $Q\Q$ and $\Q Q$, the only
nontrivial statements are the commutativity of $Q\Q$, $\Q Q$, $\m$ and $\n$. We have: 
\begin{eqnarray*}
{[\m ,\a ]}&=&{\m \a - \a \m=Q(Q\K Q-\Q Q^2 \Q )\Q }\\
&=&{Q[\a,\b]\Q =0}\\
{[\m ,\b ]}&=&{Q^2\T Q-\Q Q^3 \K }\\
&=&{Q^2(6\K H-\K \a-\b \K-Q \T)}-\\
& &-{(6Q^2 H-Q^3\Q -Q^2\Q Q-\a Q^2 )\K }\\
&=&{6\m H -\m \a -Q^2\b \K -Q^3\T-}\\
& & {-6\m H +Q^3\T +Q^2\b \K +\a \m }\\
&=&{[\a , \m]=0}
\end{eqnarray*}
In proving the second commutation relation, in the second equality we used relations $(ii)$ and $(iv)$ from the parasusy algebra. In an
analogous manner we can obtain the commutativity of $\n$ and $\a$ or $\b$. The commutativity of $\m$ and $\n$ is
trivial since we have $Q^4=Q^{\dagger 4}=0$.\kraj \\
\indent From now on, we will denote the commuting set of operators in Proposition 1 by $S$. Thus, by
restricting to any finite dimensional subspace of $\H$, the operators in $S$ can be simultaneously diagonalized.

\begin{proposition}
If $\Psi$ is an eigenstate of all the operators belonging to $S$, then so are $Q\Psi$ and $\Q \Psi$.
\end{proposition} 
\indent
\textit{Proof}: The statement is obvious for $H$ and $N$. For $\a$, the nontrivial part is for $\Q \Psi$. Since $\Psi$ is an
eigenstate of $\a$, we have $Q\Q \Psi = c \Psi$, for some scalar $c$. If $c=0$, then $\| \Q \Psi \|^2=\langle\Q \Psi,\Q \Psi\rangle = \langle \Psi,
Q\Q
\Psi\rangle =0$, i.e. $\Q \Psi = 0$ (and so $Q\Q \Q \Psi=0=\Q \Psi$). If $c \ne 0$ then we have:
$$Q \K \Psi= (1/c) Q\K Q\Q \Psi {\buildrel {(iii)} \over = }(1/c)\Q Q^2 \K \Psi=(1/c)d\Q \Psi,$$
where $d$ is the eigenvalue of $\Psi$ with respect to $\m$. 
The reasoning is analogous for $\b (Q\Psi)$. For ``higher'' cases, the nontrivial
parts are proving that $Q^2 \K \Q \Psi$ and $\n Q \Psi$ are proportional to $\Q \Psi$ and $Q \Psi$, respectively. In the first
case, we have:
\begin{eqnarray*}
\m \Q \Psi &\buildrel {(iv)}\over = & 6Q\K H\Psi -Q \K \b \Psi -Q\K \a \Psi -\a Q \K \Psi\\
&=& 6(\a)H(\Q\Psi) - (\a)\Q(\b)\Psi - (\a)(\b)\Q\Psi - (\a)^2\Q \Psi .               
\end{eqnarray*}
Now, all terms on the RHS are proportional to $\Q \Psi$ because of the first part of the proof. Again the reasoning is
analogous for $\K Q^3 \Psi$.\kraj
\\
        
\begin{definition}
Given an eigenstate $\Psi$ of $S$, the family $\F(\Psi)$ is the subspace of $\H$ spanned by $\Psi$ and all the states obtained
by operating on $\Psi$ with $Q$'s and $\Q$'s.
\end{definition}
\begin{proposition}
For all $\Psi \in \H$ which are eigenstates of $S$, we have $\dim \F(\Psi) \le 4$, and a basis of $\F(\Psi)$ is one of the
following ten sets:
\begin{eqnarray*}
&\{\Psi,\Q \Psi,\K \Psi,\T \Psi\},\{Q\Psi,\Psi,\Q \Psi,\K \Psi\},\{Q^2 \Psi,Q\Psi,\Psi,\Q\Psi\},\\
&\{Q^3\Psi,Q^2\Psi,Q\Psi,\Psi\},\{\Psi,\Q \Psi,\K \Psi\},\{Q\Psi,\Psi,\Q \Psi\},\\
&\{Q^2\Psi,Q\Psi,\Psi\},\{\Psi,\Q\Psi\},\{\Psi,Q\Psi\},\{\Psi\}.
\end{eqnarray*} 
\end{proposition}
\indent \textit{Proof}: First note that any state $\phi$ which includes both $Q$'s and $\Q$'s acting on $\Psi$ can be written
as a scalar multiple of some state with only $Q$'s or only $\Q$'s acting on $\Psi$. Indeed, we would certainly have $Q$ and
$\Q$ as neighbours somewhere, i.e. $\phi=X(Q\Q)Y\Psi$ or $\phi=X(\Q Q)Y\Psi$, where $X$ and $Y$ represent some products of $Q$'s and
$\Q$'s. But because of Proposition 2, we have that $Y\Psi$ is an eigenstate of $\a$ and so $\phi=c \cdot XY\Psi$. By
continuing this process we can eliminate either all $Q$'s or all $\Q$'s 
(whichever were less in $\phi$).
Furthermore, it is obvious that
$Q^4\Psi={\Q}^4\Psi=0, \forall \Psi \in \H$, because of the stratification of $\H$, and so we obtain the first four cases
and the fact that $\dim \F(\Psi) \le 4$. If, in addition, we have $\T \Psi=0$ or $\K \Psi=0$ or $\Q \Psi=0$ or $Q\Psi=0$ or $Q^2\Psi=0$ or
$Q^3\Psi=0$, we obtain the remaining ``degenerate'' cases.\kraj
\\

\begin{definition} 
The family $\F(\Psi)$ is called {\em generic} if $\dim \F(\Psi) = 4$.\end{definition}
Since $\Psi$ is an eigenstate of $N$ we have that all four basis vectors of a generic family belong to different subspaces $\H_j$.
\begin{proposition}
If $\Psi_i$ and $\Psi_j$ are eigenstates of $S$, and if $\Psi_i\notin \F(\Psi_j)$ then $\F(\Psi_i)\cap \F(\Psi_j)=\left\{0\right\}.$\end{proposition}
\indent \textit{Proof}: We just have to prove that two other elements
of $\F(\Psi_i)$, $Q\Psi_i$ and $\Q\Psi_i$, do not
belong to 
$\F(\Psi_j)$. Indeed, if $Q\Psi_i\ne 0$, then $\Q Q \Psi_i = c^2\Psi_i$ with $c\ne 0$ and so if $Q\Psi_i\in
\F(\Psi_j)\setminus \{0\}$
 then $\Psi_i ={1\over {c^2}}\Q \underbrace{Q\Psi_i}_{\in \F(\Psi_j)}\in \F(\Psi_j)$ which is a contradiction.\kraj \\
\\
Now, completely analogously to the cases $p=1$ and $p=2$ (see e.g. \cite{P}) we have the following theorem:
\begin{theorem} Let $(\H,H)$ be a $p=3$ parasupersymmetric pair and let the spectrum of $H$ be bounded below on $\H$ with a
finite multiplicity for each eigenvalue. Then there exists a sequence $\Psi_n \in \H$ such that\begin{center}
$\H=\bigoplus^{\infty}_{n=1} \F(\Psi_n)$
.
\end{center}\end{theorem}

\section{A generalized Witten index}
\indent The Witten index in supersymmetric quantum mechanics \cite{W} is given by $(\P_1=)\trace{(-1)}^F$ (as we said in
section 2, in this case $F=N$), and in \cite{P} it was generalized to all $p$. In particular in the 
$p=2$ parasupersymmetric case we have $\P_2=\trace{(e^{2\pi Ni/3})}$ and for $p=3$ we have the following:
\begin{definition} Given a $p=3$ parasupersymmetric pair satisfying the conditions of Theorem 1, and such that $Q$ is Fredholm, its parasupersymmetry index $\P_3$ is given by\begin{center}
$\P_3=\trace{(e^{2\pi i\over 4})}^N=\trace{i^ N}$
.
\end{center}\end{definition}
\indent As in the cases of lower $p$, we can find an expression for the parasupersymmetry index as a certain type of analytical index, in an analogous fashion to index theorems.
\begin{theorem}
If we define the analytical index of the ``paracomplex'' $C$:
$$0\buildrel{Q^{(0)}}\over\longleftarrow {\H}_0 
{{ {Q^{\dagger (0)}}\atop \longrightarrow} \atop
{\longleftarrow \atop {Q^{(1)}}}} {\H}_1 
{{ {Q^{\dagger (1)}}\atop \longrightarrow} \atop
{\longleftarrow \atop {Q^{(2)}}}} {\H}_2
{{ {Q^{\dagger (2)}}\atop \longrightarrow} \atop 
{\longleftarrow \atop {Q^{(3)}}}} {\H}_3
\buildrel{Q^{\dagger (3)}}\over\longrightarrow 0$$
%
%
where $Q^{(i)}=Q|_{ \H_i}$, $Q^{\dagger (i)}=\Q|_{ \H_i}$, to be:
\begin{eqnarray*}
\ind {}_{\rm{A}} C&=&\dim\ker{Q^{\dagger (0)}}+(1+i)\dim\ker{Q^{\dagger (1)}} + i\dim\ker{Q^{\dagger (2)}}\\
             & &-\dim\ker Q^{(1)}-(i+1)\dim\ker Q^{(2)}-i\dim\ker Q^{(3)},
\end{eqnarray*}
then:\begin{equation}
\ind {}_{\rm{A}} C=\P_3
.
\end{equation}\end{theorem}

\indent\textit{Proof:} Because of Theorem 1, $\H$ decomposes as a sum of families $\F(\Psi)$ and so $\P_3$ is a sum of traces of
$i^N$ on $\F(\Psi)$. On generic families the trace obviously vanishes since we have one basis vector in each $\H_i$ and so
$\trace i^N|_{\F(\Psi)}=i^0+i^1+i^2+i^3=0$. Thus we only have to show the equality on all types
 of nongeneric families, i.e. on the corresponding basis elements, which have $\Psi$'s from different subspaces $\H_j$. (The non-generic families are finite in number because of the Fredholm condition on $Q$.) 
This is a simple case-by-case check, for
instance, for the cases $\{\Psi_0,(\Q\Psi)_1,(\K\Psi)_2    \}$, $\{(Q\Psi)_0,\Psi_1,(\Q\Psi)_2    \}$ or $\{
(Q^2\Psi)_0,(Q\Psi)_1, \Psi_2\}$,
where the subscript $j$ denotes that the state belongs to $\H_j$, 
 we obviously have a contribution of $1+i+i^2=i$ to $\P_3$ and 1 to $\dim\ker {Q^{\dagger
(2)}}$. In other cases we work analogously. \kraj \\

\section{A model with $p=3$ parasupersymmetry}
\indent
By analogy with the $p=1$ and $p=2$ models from section 2, we can introduce a class of models with one
bosonic and three parafermionic levels, which are parasupersymmetric of order $p=3$. By a direct generalization, we can introduce
parasupersymmetry operators acting on the Hilbert space $\H=L^2(\R,{\R}^4)$ as follows:
\arraycolsep 2pt
\begin{equation}
Q\left(\begin{array}{c}f_3\\f_2\\f_1\\f_0\end{array}\right)=\left( \begin{array}{c}
0 \\
(-d_x+W_1) f_3\\
(-d_x+W_2) f_2\\
(-d_x+W_3) f_1 
\end{array} \right),\quad
\Q\left(\begin{array}{c}f_3\\f_2\\f_1\\f_0\end{array}\right)=\left( \begin{array}{c}
(d_x+W_1) f_2\\
(d_x+W_2) f_1\\
(d_x+W_3) f_0\\
0 
\end{array} \right)
\end{equation}
where the $W_i(x)$'s are superpotentials whose properties are yet to be specified. In order for the operators to satisfy 
relation $(iii)$ of our parasusy algebra, the superpotentials must satisfy:
\begin{eqnarray}
{W_2^2-W_1^2+{W_2}'+{W_1}'}&=&c \\
{W_3^2-W_2^2+{W_3}'+{W_2}'}&=&d,
\end{eqnarray} 
with $c$ and $d$ being arbitrary constants. Since we want our Hamiltonian to be of the form:
$$6H=(-3 d^2_x){\I}_4+\diag(h_1,h_2,h_3,h_4)$$
for some functions $h_i$, the condition that the parasusy operators commute with $H$ (relation $(i)$ of the parasusy algebra) is
equivalent to the following relations for the $h_i$'s:
\begin{eqnarray}
h_i&=&6W_i'+h_{i+1},\,\, i=1,2,3,
 \\ {h_1}'&=&3{({W_1}'+W_1^2)}'.
\end{eqnarray}
By using (8) and (9), we can obtain the following form for $h_1$:
\begin{equation}
h_1= W_1^2+W_2^2+W_3^2+5{W_1}'+3{W_2}'+{W_3}'+e,
\end{equation}
with $e$ being a new constant. For the other $h_i$'s we find similar expressions by using the relation (10).\\
\indent
By direct calculations, one can show that neither bilinear nor trilinear combinations of parasupercharges give appropriate
products of parasupercharges and Hamiltonians (analogously to the $p=1$ and $p=2$ parasusy algebras). However, their quadrilinear
combinations satisfy exactly the remaining relations $(ii)$ and $(iv)$ from our $p=3$ parasusy algebra, provided that
$e=0$. Indeed, in order to satisfy $(ii)$ (and thus $(iv)$ since it is the
``adjoint'' of $(ii)$), we need that $e=0$. Thus we have
obtained a model where the parasupercharges are given by (7), with the superpotentials satisfying  
\begin{eqnarray*} 
{(W_2^2-W_1^2)'+{W_2}''+{W_1}''}&=&0 \\
{(W_3^2-W_2^2)'+{W_3}''+{W_2}''}&=&0,
\end{eqnarray*} 
and with the Hamiltonian $H$ given by
\begin{equation}
H=\left( -{1\over 2} d^2_x+{1\over 6}\left( W_1^2+W_2^2+W_3^2 \right) \right)
{\I}_4 + \diag(U_1,U_2,U_3,U_4)
\end{equation}
where
\begin{eqnarray*}
U_1 & = & 5{W_1}'+3{W_2}'+{W_3}'   \\
U_2 & = & -{W_1}'+3{W_2}'+{W_3}'   \\
U_3 & = &  -{W_1}'-3{W_2}'+{W_3}'  \\
U_4 & = &  -{W_1}'-3{W_2}'-5{W_3}'
\end{eqnarray*}

The Hamiltonian is obviously diagonal and we can define $\H_0$, $\H_1$, $\H_2$ and $\H_3$ as the subspaces given by 
$\{ f_3=f_2=f_1=0\}$, $\{ f_3=f_2=f_0=0\}$, $\{ f_3=f_1=f_0=0\}$, $\{ f_2=f_1=f_0=0\}$, respectively. In this way we obtain a $p=3$
parasupersymmetric pair $(\H,H)$. The parafermion number operator in this model is given by $N=\diag(3,2,1,0)$. Note that if we multiply  $Q$ by $i$ and $\Q$ by $-i$, we obtain the physics convention form of these
operators (like in \cite{RS}). 
\\
\indent  One of the physical interpretations of this model is that, for a specific choice of superpotentials, we can obtain
a Hamiltonian which describes the one-dimensional motion of a spin $3/2$ particle in a magnetic field directed along the
third axis:
\begin{equation}
H=-\frac{1}{2} d_x^2 + V(x) + B(x)J_3,
\end{equation}
where $J_3$ is the diagonal generator of the four-dimensional representation of SO(3),
i.e. $J_3=N-3/2$. In order that the forms (13) and (14) of the Hamiltonian coincide, we must have that ${W_1}'={W_2}'={W_3}'=B(x)$
and that the superpotentials have the following forms:
\begin{eqnarray*}
W_3(x)&=&W_2(x)+\alpha=W_1(x)+2\alpha \\
W_1(x)&=&{\left\{ \begin{array}{ll}
k_1 x +k_2, & \quad \alpha=0 \\
k_1 e^{-\alpha x}+k_2, & \quad \alpha \ne 0 
\end{array} \right.}
\end{eqnarray*}
with $\alpha,k_1,k_2$ being arbitrary constants. The potential term is in all cases given by
$$V(x)=(W_1^2+W_2^2+W_3^2)/6,$$
and so we see that the case $\alpha=0$ corresponds to a harmonic oscillator in a homogeneous magnetic field, and the case
$\alpha \ne 0$ corresponds to the Morse potential and exponential magnetic field. This is in agreement with the discussion in \cite{RS}.


\section{Spectra and the index}
\indent As in \cite{RS} in the $p=2$ case, we can introduce the hermitian parasusy charges:
$$Q_1={{\Q+Q}\over {2\sqrt{3}}},\quad Q_2={{\Q-Q}\over {2\sqrt{3}i}}.$$
In terms of $Q_1$, $Q_2$ and $H$, our $p=3$ parasusy algebra from Definition 1 becomes:
\arraycolsep 2pt
\begin{eqnarray}
[Q_1, H] & = & [Q_2, H] =0 \\
\Sigma^{1,3}&=&\Sigma^{3,1}\\ \Sigma^{4,0}+\Sigma^{0,4}&=&\Sigma^{2,2}\\
\Sigma^{1,3}+\Sigma^{3,1}&=&\Sigma^{1,1}H\\
2(Q_1^4-Q_2^4)&=&(Q_1^2-Q_2^2)H\\
Q_1^3Q_2+Q_2^2Q_1Q_2+Q_2Q_1Q_2^2+Q_2Q_1^3&=&Q_2^3Q_1+Q_1^2Q_2Q_1+Q_1Q_2Q_1^2+Q_1Q_2^3, 
\end{eqnarray}
\arraycolsep 6pt
where $\Sigma^{i,j}$ denotes the sum of all monomials with $i$ $Q_1$'s and $j$ $Q_2$'s. 
We may use this form of the parasusy algebra to show that the eigenvalues of $H$ are generically non-negative. Since $Q_1$ and $H$ commute we can diagonalize them simultaneously. Let $\Psi$ be a simultaneous eigenstate of  $Q_1$ and $H$ with eigenvalues $q_1$ and $E$ respectively. Then from equation (19) we have the equation
\[
2(q_1^2 + Q_2^2) (q_1^2 - Q_2^2)\Psi = 2(q_1^4 - Q_2^4)\Psi= 
(q_1^2 - Q_2^2)E\Psi.
\]
Thus if $(q_1^2 - Q_2^2)\Psi \neq 0$ we deduce that $E$ is non-negative, since it is an eigenvalue of the positive operator $2(q_1^2 + Q_2^2)$. Therefore the only states which may have negative energy eigenvalues are those belonging to 
${\rm ker}(Q_2^2 - q_1^2)$. Let us assume that $\Psi$ belongs to this kernel. Furthermore, from (18) we obtain:
\[
(Q_1^2+Q_2^2)(Q_1Q_2+Q_2Q_1)\Psi + (Q_1Q_2+Q_2Q_1)(Q_1^2+Q_2^2)\Psi =(Q_1Q_2+Q_2Q_1)E\Psi
\]
which gives:
\begin{equation}
(Q_1^2+Q_2^2)(Q_1Q_2+Q_2Q_1)\Psi =(E-2q_1^2)(Q_1Q_2+Q_2Q_1)\Psi. \end{equation}
So, if $(Q_1Q_2+Q_2Q_1)\Psi \ne 0$ then $E-2q_1^2$ (and hence $E$) must be nonnegative since it is
an eigenvalue of the positive operator $(Q_1^2+Q_2^2)$. So, suppose that $(Q_1Q_2+Q_2Q_1)\Psi = 0$.
In this case, we see that all relations (15)-(20) of our parasusy algebra are satisfied.
Now if we go back to $Q$ and $\Q$, we obviously have:
\begin{equation}
Q^2 \Psi= 3{(Q_1- iQ_2)}^2\Psi=3(Q_1^2-Q_2^2)\Psi-3i(Q_1Q_2+Q_2Q_1)\Psi=0,
\end{equation}
and analogously 
\begin{equation}
\K \Psi=0. 
\end{equation}
Furthermore, since $\Psi$ is an eigenfunction of $Q_1$, we have 
\begin{equation} Q\Psi +\Q\Psi =c\Psi,\end{equation} 
for some real $c$. We conclude that the state $\Psi$ can have negative energy, if and only if it satisfies (22), (23) and (24). 
Thus, unlike in the $p=1$ supersymmetric case, where the energy spectrum is
positive definite (see \cite{W}), there is no restriction on the parasupersymmetric vacuum energy 
in the $p=3$ case (in the same way as in the $p=2$ case - see \cite{RS}).\\
\indent Because of the conservation of parasupercharges, and because of the 4-fold stratification of $\H$, we know that the
spectrum of
$H$ is generically 4-fold degenerate, except for a few lowest energy levels. Indeed, since $N$ and $H$ commute, we can diagonalize them simultaneously. Then because of $\left[N , \Q \right]=\Q$, $Q^{\dagger 4}=0$ and since 
$\left[ \Q ,H \right]=0$, any state $\phi$ with zero parafermions ($N=0$), and the states $\Q \phi$, $\K \phi$ and 
$\T \phi$, have the equal energy (eigenvalue of Hamiltonian).  
So, obviously, in the generic case, for each eigenvalue, we would have four linearly independent eigenstates (one from each
$\H_j$). The only way to have less then 4-fold degeneracy is when the family generated by $\Psi$ (i.e. $\F(\Psi)$)
is nongeneric. Thus, we have to look at $\ker Q$ and $\ker \Q$.\\
\indent Our goal is to connect the possible forms of the spectrum of $H$ from our class of models from Section 5, with the corresponding index. By using theorem 2 (i.e.
formula (6)), we see that for calculating the index we need to know the corresponding dimensions of the kernels of the restrictions of the 
parasupercharges on different $\H_i$'s. For instance, the only way for $\dim \ker Q^{(3)}$ to be non-zero is when
the vector ${(\Psi^{(1)}_+,0,0,0)}^{\mathrm{T}}$ is normalizable in $L^2(\R,\R^4)$, where
$$ \Psi^{(i)}_{\pm}=\exp\left( \pm \int^x{W_i(t)dt}\right),\quad i=1,2,3.$$
When this vector is normalizable, we obviously have that the dimension of the kernel of the corresponding operator is
equal to 1. Thus we have:
\begin{eqnarray*}
\dim \ker Q^{(3)} = 1 & \Longleftrightarrow & \Psi^{(1)}_+ \in L^2(\R,\R) \\
\dim \ker Q^{\dagger (2)} = 1 & \Longleftrightarrow & \Psi^{(1)}_- \in L^2(\R,\R) \\
\dim \ker Q^{(2)} = 1 & \Longleftrightarrow & \Psi^{(2)}_+ \in L^2(\R,\R) \\
\dim \ker Q^{\dagger (1)} = 1 & \Longleftrightarrow & \Psi^{(2)}_- \in L^2(\R,\R) \\
\dim \ker Q^{(1)} = 1 & \Longleftrightarrow & \Psi^{(3)}_+ \in L^2(\R,\R) \\
\dim \ker Q^{\dagger (0)} = 1 & \Longleftrightarrow & \Psi^{(3)}_- \in L^2(\R,\R), 
\end{eqnarray*}
with the dimensions of the kernels being equal to 0 in all other cases.\\
\indent Thus the spectrum of $H$ and the index for our model depend on the possible nontrivial kernels of the 
parasupercharges (and their restrictions), which in turn depend on the square integrability of the functions
$\Psi^{(i)}_{\pm}$. Since, for each $i$ we have that either $\Psi^{(i)}_+$ is normalizable or
$\Psi^{(i)}_-$ is normalizable or neither of them is, we can theoretically have $3^3=27$ different outcomes for the spectrum and the
value of the index. One can calculate the value of the index and sketch the form of the spectrum 
in all possible cases. One can also find explicit examples for $W_i$ corresponding to all these cases. In the cases of lower
$p$, by repeating the same procedure, one can obtain a one-to-one correspondence between 
the spectrum and the index (see \cite{RS} for an analysis of the spectrum in $p=2$ case). \\
\indent However in the $p=3$ case there are only 19 different outcomes for the value of the index, so the forms of the
spectrum will also depend on one more parameter - namely the sign of the constants $c$ and $d$ from equations (8) and (9).
Furthermore, the constant $c$, together with one more constant $c_0$ and a function $K$ classify all possible solutions of 
equation (5) (identical to equation (8)), which is the only restriction in the 
$p=2$ case (\cite{RS}, \cite{P}). Rewriting equation (8) as follows:
$$(W_1+W_2)'+(W_1+W_2)(W_1-W_2)=c,$$
it becomes a linear differential equation for the function $W_1+W_2$. If we denote
$$K=\int^x {(W_2 - W_1)},$$
its solution is given by the formula:
$$W_1+W_2=e^{-K} \left( c_0+c \int_0^{x} {e^K} \right),\quad c_0 \in \R,$$
which together with $W_1-W_2=-K'$, gives:
\begin{equation}
W_{1,2}={1 \over 2} \left( \mp K'+c_0 e^{-K}+ce^{-K} \int_0^{x} {e^K} \right) .
\end{equation}
Thus, in the case $p=2$, the class of models is characterized by one function ($K$) and two constants ($c,c_0$). In the $p=3$ case we have
to go on and solve equation (9), where now $W_2$ is a known function (depending on $K,c,c_0$), and so we just have a first-order equation (the Riccati equation) with respect to $W_3$. This will give us one new constant $d_0$, which, together with the existing
$d$, shows that our class of models is characterized by one function and four constants. \\
\\
{\bf  {Acknowledgments}} \quad This work was supported by 
{\it{Programa Operacional ``Ci\^{e}ncia, Tecnologia, Inova\c{c}\~ ao''}}
(POCTI) of the {\it{Funda\c{c}\~{a}o para a Ci\^{e}ncia e a Tecnologia}} 
(FCT), cofinanced by the European Community fund FEDER. The first author 
(Marko Sto\v si\'c) is supported by the 
{\it{Funda\c{c}\~{a}o para a Ci\^{e}ncia e a Tecnologia}} (FCT), 
grant no. SFRH/BD/6783/2001.


\footnotesize
 \begin{thebibliography}{99}
\bibitem{BD}
 Beckers, J. and Debergh, N.: ``On parasupersymmetry and remarkable Lie structures'', J.Phys A 23, no. 14, L751--L755 (1990).
\bibitem{DMSV}
 Durand, S., Mayrand, M., Spiridonov, V. and Vinet, L.: ``Higher order parasupersymmetric quantum mechanics'', Mod. Phys. Lett.
A, Vol. 6, no. 34 3163-3169 (1991).
\bibitem{GK} Gendenshtein, L. E. and Krive, I. V.: ``Supersymmetry in quantum mechanics'', Soviet Phys. Uspekhi 28 (1985), no. 8, 645--666 (1986); translated from Uspekhi Fiz. Nauk 146, no. 4, 553--590 (1985) (Russian).
\bibitem{G} Green, H. S.: ``A generalized method of field quantization'', Phys. Rev. (2) 90 270-273 (1953).
\bibitem{OK}
 Ohnuki, Y. and Kamefuchi, S.: ``Quantum Field Theory and Parastatistics''. University of Tokyo Press, Springer Verlag (1982).
\bibitem{P}
 Picken, R.: ``An index from parasupersymmetry'', preprint (1990).
\bibitem{RS}
 Rubakov, V.A. and Spiridonov, V.P.: ``Parasupersymmetric quantum mechanics'', Modern Physics Letters A3, 1337-1347 (1988).
\bibitem{W}
 Witten, E.: ``Dynamical breaking of supersymmetry'', Nucl. Phys. B185, 513-554 
(1981), and 
 ``Constraints on supersymmetry breaking'', Nucl. Phys. B202, 253-316 (1982).
 
\end{thebibliography}
\end{document}